\documentclass[11pt]{article}

\usepackage{amsmath,amssymb,amsfonts,latexsym,epsfig, varioref,setspace,verbatim}

\usepackage{times}
\usepackage{bm}
\usepackage{natbib}
\usepackage{multirow}
\usepackage{mathtools}
\usepackage{mathrsfs}
\usepackage{xcolor}
\usepackage{soul}

\usepackage[plain,noend]{algorithm2e}

\newcommand{\Bb}{\beta}
\newcommand{\Ba}{ \alpha}
\newcommand{\Bg}{\gamma}
\newcommand{\Btau}{\tau}
\newcommand{\Bt}{\theta}

\newcommand{\Bl}{\lambda}

\newcommand{\BZ}{{Z}}
\newcommand{\BX}{{X}}
\newcommand{\BB}{{B}}
\newcommand{\BD}{{D}}

\newcommand{\BW}{{W}}

\newcommand{\BS}{{S}}

\newcommand{\BM}{{M}}

\newcommand{\wh}{\widehat}

\newcommand{\tg}{\textsl{g}}
\newcommand{\tj}{\mathcal{J}}
\DeclareMathOperator*{\argmin}{\arg\!\min}
\DeclareMathOperator*{\argmax}{\arg\!\max}

\newtheorem{condition}{Condition}
\newtheorem{theorem}{Theorem}
\newtheorem{remark}{Remark}

\newcommand{\blind}{1}

\begin{document}

\def\spacingset#1{\renewcommand{\baselinestretch}%
	{#1}\small\normalsize} \spacingset{1}


\if1\blind
{
	\title{\bf A penalized likelihood approach for efficiently estimating a partially linear additive transformation model with current status data}
	\author{Yan Liu, Minggen Lu \\          
		School of Community Health Sciences, University of Nevada, Reno\\ \hspace{.2cm}\\
		Christopher S. McMahan\\           
		Department of Mathematical Sciences, Clemson University\\
		}
	\maketitle
} \fi

\if0\blind
{
	\bigskip
	\bigskip
	\bigskip
	\begin{center}
		{\LARGE\bf A penalized likelihood approach for efficiently estimating a partially linear additive transformation model with current status data}
	\end{center}
	\medskip
} \fi


\begin{abstract}
	\noindent Current status data are commonly encountered in medical and epidemiological studies in which the failure time for study units is the outcome variable of interest. Data of this form are characterized by the fact that the failure time is not directly observed but rather is known relative to an observation time; i.e., the failure times are either left- or right-censored. Due to its structure, the analysis of such data can be challenging. To circumvent these challenges and to provide for a flexible modeling construct which can be used to analyze current status data, herein, a partially linear additive transformation model is proposed. In the formulation of this model, constrained $B$-splines are employed to model the monotone transformation function and nonlinear covariate effects. To provide for more efficient estimates, a penalization technique is used to regularize the estimation of all unknown functions. An easy to implement hybrid algorithm is developed for model fitting and a simple estimator of the large-sample variance-covariance matrix is proposed. It is shown theoretically that the proposed estimators of the finite-dimensional regression coefficients are root-$n$ consistent, asymptotically normal, and achieve the semi-parametric information bound while the estimators of the nonparametric components attain the optimal rate of convergence. The finite-sample performance of the proposed methodology is evaluated through extensive numerical studies and is further demonstrated through the analysis of uterine leiomyomata data. 

\end{abstract}

\noindent%

\vfill

\newpage

\section{Introduction}
\noindent Current status data is characterized by the fact that the failure/event time cannot be directly measured, but rather is known relative to a single observation time. That is, the failure time is known to be smaller or larger than an observation time leading to left- or right-censored data, respectively. Data of this structure commonly arise as a part of medical and epidemiological studies, among many others, in which study participants are observed only once due to destructive/invasive testing, resource limitations, etc. For example, as part of the Right From The Start (RFTS) study, participants completed an ultrasound examination as a means to determine their fibroid status \citep{laughlin2009prevalence}. In this study, fibroid onset time could not be directly observed but was known relative to the participant's age at the ultrasound examination time.

For the regression analysis of current status data, many parametric and semi-parametric methods have been proposed. In particular, many of these efforts have focused on the proportional hazards \citep{david1972regression} and proportional odds models \citep{dabrowska1988estimation}; for a review see  \cite{mcmahan2013regression}, \cite{lu2017penalized}, and the references therein. To provide a more general modeling framework, \cite{gu2005baseline}, \cite{sun2005semiparametric}, \cite{zhang2005regression}, \cite{zhang2013empirical}, and \cite{zeng2016maximum} developed estimation techniques under linear transformation models. A potentially prohibitive assumption made in all of the aforementioned works is that the covariate effects on the failure time are ``linear." To relax this assumption, \cite{cheng2011semiparametric} developed a semi-parametric additive transformation model and \cite{lu2018partially} presented a partially linear PH model for current status data. For a comprehensive review of methods to analyze interval-censored data see \cite{sun2007statistical} and \cite{banerjee2012current}.

In this article, we seek to develop a novel methodology which can be used to conduct the regression analysis of current status data. In particular, a very general partially linear additive transformation model is proposed. The proposed model makes use of constrained $B$-splines to approximate all unknown functions; i.e., the transformation and regression functions. In order to provide for more efficient estimators, a penalization strategy is utilized to regularize the estimation of the unknown functions. To complete model fitting a computationally efficient and easy to implement hybrid algorithm is developed. Theoretically, it is shown that the penalized estimators of the regressions coefficients are asymptotically normal and efficient; i.e., they attain the semi-parametric information bound. Moreover, we establish that the estimators of the nonparametric components achieve the optimal rate of convergence for an appropriately chosen order of the smoothing parameters. Finite sample techniques of determining the smoothing parameter and estimating the asymptotic covariance matrix of the estimated regression coefficients are provided and evaluated.   

\section{Model}

\noindent Consider a study which examines $n$ subjects for a failure time of interest. Let $T_i$ denote the time that the $i$th subject experiences the failure, for $i=1,....,n$. To provide for modeling flexibility, we assume that $T_i$ obeys a partial linear additive transformation model which is given by
\begin{equation}\label{model1}
\eta(T_i) = -\BZ^\intercal_i\Bb-\sum_{j=1}^J \varphi_j(W_{ij})+\varepsilon_i,
\end{equation}
where $\eta(\cdot)$ is an unknown monotone increasing transformation function on $(0,\infty)$ with $\eta(0) = -\infty$ and $\varepsilon_i$ is a random variable having cumulative distribution function (CDF) $H(\cdot)$. In the above specification, $\Bb$ is a $q$-dimensional vector of regression coefficients corresponding to the covariates $\BZ_i=(Z_{i1},...,Z_{iq})^\intercal$ and  $\varphi_j(\cdot)$ is an unknown smooth regression function taking the covariate $W_{ij}$ as its argument, for $j=1,...,J$. Throughout, we assume that $\varepsilon_i \perp \varepsilon_j$, for $i\neq j$, and $\varepsilon_i \perp \BX_i$, where $\perp$ denotes statistical independence and $\BX_i = (\BZ_i^\intercal, W_{i1},...,W_{iJ})^\intercal$. This formulation provides for a very general modeling construct that holds several popular models as special cases. For example, specifying $H(\cdot)$ to be the CDF of the extreme value distribution causes \eqref{model1} to reduce to the standard proportional hazards (PH) model, while taking $H(\cdot)$ to be the CDF of the standard logistic distribution results in the proportional odds (PO) model. 

Following the work of  \cite{murphy1997maximum}, \cite{sun2005semiparametric}, and \cite{cheng2011semiparametric}, we note that \eqref{model1} can be rewritten in terms of the conditional CDF of $T_i$, given the covariate vector $\BX_i$, and can be expressed as 
\begin{align}\label{model}
F(t\mid \BX_i) =\tg^{-1}\left\{\BZ^\intercal_i\Bb+ \eta(t) + \sum_{j=1}^J \varphi_j(W_{ij})\right\},
\end{align}
where $\tg(\cdot)$ is a known, smooth, and strictly increasing link function on $(0,1)$. Though equivalent to \eqref{model1}, the representation provided in \eqref{model} is more convenient for reasons that will shortly become apparent. Here $\tg(\cdot)$ assumes the role of $H(\cdot)$ with respect to determining the final form of the proposed model; e.g., specifying $\tg(x) = \log\{-\log(1-x)\}$ provides for the PH model and $\tg(x) = \log\{x/(1-x)\}$ results in the PO model.

A key feature of current status data is that the failure times (i.e., $T_i$, for $i=1,\dots,n$) are not directly observed but rather are known relative to observation times; i.e., the $T_i$ is either left- or right-censored indicating that $T_i$ occurs before or after the observation time, respectively. To mathematically frame the structure of this data, let $Y_i$ denote the observation time for the $i$th individual and let $\Delta_i = I(T_i < Y_i)$ denote the corresponding censoring indicator, where $I(\cdot)$ is the usual indicator function. Note, $\Delta_i=1(0)$ represents the event that the $i$th individual's failure time is left(right)-censored. Thus, the observed data consist of $\{(Y_i,\Delta_i,\BX_i)\}_{i=1}^n$. Following the works of \cite{mammen1997penalized}, \cite{ma2005penalized}, and \cite{ma2009cure}, we propose to estimate the unknown parameters via the following penalized log-likelihood
\begin{align*} 
l_{\Bl}(\tau) =\sum_{i=1}^n[\Delta_i\log F(Y_i\mid \BX_i)+&(1-\Delta_i)\log\{1-F(Y_i\mid \BX_i)\}]
-\frac{\lambda_0^2}{2}\tj^2(\eta)-\sum_{j=1}^J\frac{\lambda_j^2}{2}\tj^2(\varphi_j),
\end{align*}
where $\tau=(\beta,\eta,\varphi_1,...,\varphi_J)$ represents the collection of unknown model parameters, $\Bl = (\lambda_0,\ldots,\lambda_J)^\intercal$ is a vector of smoothing  parameters, such that $\lambda_j \geq 0$ for all $j$, and
\begin{equation*}
\tj^2(\eta)=\int \{\eta^{(r)}(t)\}^2 dt,\quad \tj^2(\varphi_j)=\int \{\varphi_j^{(r)}(W)\}^2 dW.
\end{equation*}
In the penalties above, $\eta^{(r)}$ and $\varphi_j^{(r)}$ denote the $r$th derivative of $\eta(\cdot)$ and $\varphi_j(\cdot)$, respectively, $r\ge 1$. This penalized log-likelihood is derived under the assumption that the covariates are time-invariant and that the failure and censoring times are conditionally independent, given the covariates. These assumptions are common among the literature; e.g., see \cite{ma2005penalized}, \cite{cheng2011semiparametric}, \cite{banerjee2012current}, and the references therein.

\section{Large sample properties} \label{asymptotic}

Let $\tau_0=(\beta_0,\eta_0,\varphi_{10},...,\varphi_{J0})$ and $\widehat{\tau}=(\wh{\beta},\wh{\eta},\wh{\varphi}_1,...,\widehat{\varphi}_J)$ denote the true value  and the penalized maximum likelihood estimator of $\Btau$, respectively. We established the asymptotic properties of the penalized estimators $\wh{\Btau}$ under the following regularity conditions:

\begin{condition}
	The true parameter $\Bb_0$ is an interior point of a compact subset of $\mathbb{R}^q$.
\end{condition}

\begin{condition}
	The $r$th derivatives of $\eta_0$ and $\varphi_{0j}$, for $1 \le j \le J$, satisfy the Lipchitz condition on $[c,d]$ for $c > 0$ and $[a_j,b_j]$, respectively, and $\eta_0$ is strictly increasing on $[c,d]$.
\end{condition}

\begin{condition}
	The support of the observation times is an interval within $[l_y,u_y]$ with $0< l_y<u_y<\inf\{t: F_0(t) = 1\}$, where $F_0$ is the CDF of the observation times. 
\end{condition}

\begin{condition}
	The covariate  vector $\BZ$ has bounded support.
\end{condition}

\begin{condition}
	The first derivative of the link function $\tg$ is bounded away from 0 and the second derivative of $\tg$ is bounded on (0,1).
\end{condition}

\begin{condition}
	The joint density of $(Y,\BZ,\BW,\Delta)$ is bounded away from 0 and $\infty$. 
\end{condition}

\begin{condition}
	For any $\Bb\neq\Bb_0$, $\Pr(\BZ^\intercal\Bb\neq\BZ^\intercal\Bb_0)>0$ and $E\{\varphi_j(W_j)\}= 0$, ~ for $1 \le j \le J$. 
\end{condition}

\begin{condition}
	The smoothing parameters are specified such that
	$$ 
	\lambda_j^2 = o_p(n^{-1/2})~~~\mbox{and}~~~\lambda_j^{-1} =
	O_p(n^{r/(2r+1)}),~\mbox{for}~ 1 \le j \le J. 
	$$ 
\end{condition}

\begin{condition}
	The efficient information $\mathscr{I}(\Bb_0)$ defined in Appendix 1 of the Supplementary Material is non-singular.
\end{condition}


\begin{remark}
	Condition 1 is a standard assumption in semi-parametric estimation. The smoothness assumption of $\eta_0$ and $\varphi_{0j}$ in condition 2 guarantees that the unknown nonparametric components can be well approximated by $B$-splines.  Conditions 3--6 are necessary in entropy calculation used to derive the rate of convergence of $\wh{\Btau}$ and the asymptotic normality of $\wh{\Bb}$. Condition 7 is required to establish the identifiability of the proposed semi-parametric model. Condition 8 is a typical assumption used to derive the asymptotic property of penalized estimators. Finally, condition 9 is common for the asymptotic normality and the efficiency of $\wh{\Bb}$.
\end{remark}

Using modern empirical process theory, we establish the following asymptotic properties: 
\begin{theorem} (Consistency and rate of convergence)~
	\label{convergence} Under conditions 1--8, the penalized estimator $\wh\Bb$ is asymptotically consistent for $\Bb_0$, $\|\wh\eta\|_\infty = \|\wh\varphi_j\|_\infty = O_p(1)$, and $\|\widehat\eta-\eta_0\|_2= \|\widehat\varphi_j-\varphi_{j0}\|_2=O_p(n^{-r/(1+2r)})$, for $1\le j\le J$. 
\end{theorem}

\begin{theorem} (Asymptotic normality and efficiency)\label{Normality} Under conditions 1--9, 
	$$
	n^{1/2}(\wh\Bb-\Bb_0) 
	\overset{d}{\longrightarrow} \mathcal{N}\{0,\mathscr{I}^{-1}(\Bb_0)\},~\mbox{as}~n\rightarrow\infty.
	$$
\end{theorem}

The proofs of these results are relegated to Appendix 1 of the Supplementary Material. It is worthwhile to point out that if the smoothing parameters are chosen to be of the right order (see condition 8),  our functional estimates attain the optimal rate of convergence; i.e., $O_p(n^{r/(1+2r)})$. Moreover, $\wh\Bb$ is efficient in the semi-parametric sense since it achieves the information bound.

\section{Estimation}
\subsection{Penalized spline log-likelihood}

In general, estimating $\Bb$, $\eta$ and $\varphi_j$ directly from the penalized log-likelihood can be a tumultuous task. Thus, motivated by \cite{xiang1997approximate}, \cite{ma2005penalized}, and \cite{lu2017penalized}, we propose to strike a balance between modeling flexibility and complexity by approximating all unknown functions in \eqref{model} through the use of constrained $B$-splines functions. In particular,
\begin{equation*}
\eta(\cdot) \approx \eta_n(\cdot~\mid \gamma) = \sum_{k=1}^{p_0}\gamma_k b_k(\cdot) ~\textrm{ and }~ \varphi_j(\cdot) \approx \varphi_{nj}(\cdot~\mid \alpha_j^*) = \sum_{k=1}^{p_{j}}\alpha_{jk}^*b_{jk}(\cdot),
\end{equation*}
where $b_{k}(\cdot)$ and $b_{jk}(\cdot)$ are $B$-spline basis functions with unknown coefficients $\Bg =(\gamma_1,\ldots,\gamma_{p_0})^\intercal$ and $\alpha_j^*=(\alpha_{j1}^*,...,\alpha_{jp_{j}}^*)^\intercal$, respectively, $1\le j\le J$. Once a knot set is chosen and the degree of the spline function is specified these basis functions are uniquely determined; for further discussion see \cite{schumaker2007spline}. Constraints are placed on the spline coefficients to insure the monotonicity of the transformation function and the identifiability of the regression functions. The monotonicity of $\eta_n(\cdot~|\gamma)$ can be insured by requiring that $\gamma_1\le \ldots\le\gamma_{p_0}$; see Theorem 5.9 of \cite{schumaker2007spline}. As with all additive models, $\varphi_{j}(\cdot)$ is only identifiable up to an additive constant. Thus, sum-to-zero constraints are imposed such that $\sum_{i=1}^n\varphi_{nj}(W_{ji}\mid \alpha_j^*) = 0$, for $1\le j \le J$.

For implementation purposes, the sum-to-zero constraints can be expressed as ${B_j^*}^\intercal\alpha_j^*=0$, where $B_j^* = \sum_{i=1}^n B_{ji}^*$, $B_{ji}^* = \{b_{j1}(W_{ji}),\ldots,b_{jp_{j}}(W_{ji})\}^\intercal$, and $\alpha_j^*=(\alpha_{j1}^*,...,\alpha_{jp_{j}}^*)^\intercal$. Motivated by \cite{wood2017generalized}, the model is reparameterized such that the constrained set of parameters $\alpha_j^*$ can be uniquely determined by an unconstrained reduced set $\alpha_j=(\alpha_{j1},...,\alpha_{jp_{j}-1})^\intercal$. In particular, let $Q_j$ denote a $p_{j}\times(p_{j}-1)$ semi-orthogonal matrix whose columns are orthogonal to $B_j^*$. Obviously, ${B_j^*}^\intercal Q_j \alpha_j = 0$, for all $\alpha_j\in \mathbb{R}^{p_{j}-1}$, and for any $\alpha_j^*$ satisfying the sum-to-zero constraint there exists $\alpha_j$ such that $\alpha_j^*=Q_j \alpha_j $. Note, $Q_j$ can easily be found via a QR-decomposition of $B_j^*$. Thus, the log-likelihood under the proposed spline model is obtained by replacing $F(Y_i\mid X_i)$ in the penalized log-likelihood function by
\begin{equation}\label{smodel}
F_n(Y_i\mid \BX_i) =\tg^{-1}\left\{ \BZ_i^\intercal\Bb+ \BM_i^\intercal\Bg+\sum_{j=1}^J \BB_{ji}^\intercal\Ba_j\right\},
\end{equation}
where $\BM_i = \{b_1(Y_i),\ldots,b_{p_0}(Y_i)\}^\intercal$ and $\BB_{ji} = Q_{j}^\intercal B_{ji}^*$, for $1\le i\le n$. Thus, the set of unknown parameters to be estimated are given by $\Bt=(\Bb^\intercal,\Bg^\intercal, \Ba^\intercal)^\intercal$, where $\Ba=(\Ba_1^\intercal,\ldots,\Ba_J^\intercal)^\intercal$. 

To further alleviate computational burdens, we follow the proposal of \cite{eilers1996flexible} and replace the penalty terms $\mathcal{J}^2(\eta)$ and $\mathcal{J}^2(\varphi_j)$ by discrete approximations. Under this strategy, the penalized spline log-likelihood function is given by
\begin{align}\label{penalized}
l_{n,\Bl}(\theta) =\sum_{i=1}^n[\Delta_i\log F_n(Y_i\mid \BX_i)&+(1-\Delta_i)\log\{1-F_n(Y_i\mid \BX_i)\}] \nonumber \\
&-\frac{\lambda_0^2}{2} \|\BD_0^{(r)}\Bg\|^2 - \sum_{j=1}^J\frac{\lambda_j^2}{2}\|\BD_j^{(r)}\Ba_j\|^2,
\end{align}
where $\BD_j^{(r)}$ is the discrete difference operator of order $r$ \citep{eilers1996flexible,tibshirani2014adaptive} and $\|\cdot\|$ is the Euclidean norm. This operator is best defined recursively as $\BD_j^{(r+1)}=\BD_j^{(1,r)}\BD_j^{(r)}$, where 
\begin{align*}
\BD_j^{(1,r)} =\left[ \begin{tabular}{rrrrrr} $-1$ & $1$ & $0$ & $\ldots$ & $0$ & $0$\\ 
$0$ &$-1$ & $1$ & $\ldots$ & $0$ & $0$\\
$\vdots$&$\vdots$&$\vdots$& &$\vdots$&$\vdots$\\
$0$ & $0$ & $0$ & $\ldots$ & $-1$ & $1$\end{tabular}      \right] \in \mathbb{R}^{(p_j-r-2)\times (p_j-r-1)},  
\end{align*}
where $\BD_j^{(1)}=\BD_j^{(1,0)}$.
Proceeding in this fashion for $r=1$ and $2$ yields penalties of the form 
\begin{align*}
\|\BD_0^{(1)}\Bg\|^2&=\sum_{k=2}^{p_0}(\gamma_k - \gamma_{k-1})^2, \\
\|\BD_0^{(2)}\Bg\|^2&=\sum_{k=3}^{p_0}(\gamma_k - 2\gamma_{k-1}+\gamma_{k-2})^2,   
\end{align*}
respectively. Through extensive numerical studies, see Section 5, we have found that setting $r=2$ is sufficient when implementing the proposed approach.


Due to these approximations, the proposed penalized spline estimator is different from the penalized estimator, but as long as the number of basis functions for each of the unknown functions grows at least at the rate $O(n^{1/5})$, this difference will vanish and the statistical performance of the two estimators should be indistinguishable; for further discussion see \cite{eilers1996flexible}, \cite{ma2005penalized}, \cite{wood2017generalized}, and the references therein. That is, the penalized spline estimator should inherit the theoretical properties of the penalized maximum likelihood estimator. This assertion is supported by the results of the numerical experiments reported in Section 5. To ensure the accuracy of the approximation, it is generally advised that the number of knots for each function be chosen in a neighborhood of $n^{1/3}$; for further discussion see Section 5. 

\subsection{Parameter estimation}
To complete model fitting a computationally efficient hybrid algorithm is developed which can be used to identify both the penalized spline estimators and the smoothing parameters via a nested iterative process. The inner process is aimed at identifying the value of $\Bt$ that maximizes (\ref{penalized}) for a fixed value of $\Bl$, while adhering to the monotonicity constraints; i.e., the inner step identifies $\widetilde{\Bt}_{\Bl}=\argmax_{\Bt}l_{n,\Bl}(\Bt)$ s.t. $\gamma_1 \leq \ldots \leq \gamma_{p_0}$. To solve this constrained optimization problem, an iterative algorithm is developed. In each iteration of this algorithm, the unconstrained updates of the parameters, say $\Bt^{*}$, are first determined based on the current parameter values, say $\Bt^{(m)}$, via the following Fisher-scoring step 
\begin{align*}\label{fisher-score}
\Bt^{*} = \Bt^{(m)} + \mathcal{I}^{-1}_{n,\Bl}(\Bt^{(m)})\nabla l_{n,\lambda}(\Bt^{(m)}),
\end{align*}
where $\nabla l_{n,\lambda}(\Bt^{(m)})$ and $\mathcal{I}_{n,\Bl}(\Bt^{(m)})$ are the gradient and the expected value of the negative hessian matrix of (\ref{penalized}) evaluated at $\Bt^{(m)}$, respectively. Closed form expressions for these quantities are given by
\begin{eqnarray*}
	\nabla l_{n,\lambda}(\Bt)& = &\mathcal{X}^\intercal\Omega(\theta) \Delta(\theta)-S\Bt \\
	\mathcal{I}_{n,\Bl}(\Bt) &=& 
	\mathcal{X}^\intercal\Omega(\theta) \mathcal{X} +S,
\end{eqnarray*}
where $\mathcal{X}=(Z,M,B_1,...,B_J)$, $\BZ^\intercal = (\BZ_1,\ldots,\BZ_n)$, $\BM^\intercal = (\BM_1,\ldots,\BM_n)$, and $\BB_j^\intercal = (\BB_{j1},\ldots,\BB_{jn})$, and $\BS=\sum_{j=0}^J {\lambda_j^2} \BS_j $, with $S_j=\textrm{diag}\{0_{r_{1j}\times r_{1j}},\BD_j^{(r)^\intercal}\BD_j^{(r)},0_{r_{2j}\times r_{2j}}\}$, $ r_{1j}=q+p_0+...+p_{j-1}-(j-1)$, and $r_{2j}=p_{j+1}+...+p_{J}-(J-j)$. In the expressions above, the transformed response vector and weight matrix for a specific value of $\theta$ are given by   \begin{eqnarray*} 
	\Delta(\theta) & = & \left\{ \frac{\Delta_1-\pi_1}{G'(\mathcal{X}_1^\intercal \theta)},...,\frac{\Delta_n-\pi_n}{G'(\mathcal{X}_n^\intercal \theta)} \right\}^\intercal   \\
	\Omega(\theta) & = & \textrm{diag} \left\{ \frac{G'(\mathcal{X}_1^\intercal \theta)^2}{\pi_1(1-\pi_1)},...,\frac{G'(\mathcal{X}_n^\intercal \theta)^2}{\pi_n(1-\pi_n)}   \right\},
\end{eqnarray*}
where $\mathcal{X}_i$ is the $i$th row of $\mathcal{X}$, $G'(\cdot)$ is the derivative of $G(\cdot)=\tg^{-1}(\cdot)$, and $\pi_i=\tg^{-1}(\mathcal{X}_i^\intercal \theta)$.

The updated values of $\gamma$ available in $\Bt^{*}$ do not necessarily satisfy the monotonicity constraints; see Section 2. To enforce these conditions, isotonic regression is implemented to project the unconstrained estimates into the constrained space; i.e., into $\mathscr{G} = \{\Bg: \gamma_1\le\ldots\le\gamma_{p_0}\}$. This is accomplished by solving the following minimization problem 
\begin{equation*}\label{projection}
\Bg^{(m+1)} = \argmin_{\Bg\in\mathscr{G}} \left\{\sum_{k=1}^{p_0}\sigma_{k}^2(\gamma_k-\gamma_k^*)^2\right\},
\end{equation*}
where $\sigma_{k}^2$ is the diagonal element of $\mathcal{I}^{-1}_{n,\Bl}(\Bt^*)$ corresponding to $\Bg_k$, for $k=1,...,p_0$. This problem can be solved using the \textsf{pava} function in the \textsf{R} package \textsf{Iso}, which takes $\Bg^*$  and $(\sigma_{1}^2,\ldots,\sigma_{p_0}^2)^\intercal$ as inputs. Substituting  $\Bg^{(m+1)}$ for $\Bg^*$ in $\Bt^*$ one obtains the constrained update $\Bt^{(m+1)}$. These two steps are completed in turn until convergence is attained; i.e., the difference between $\Bt^{(m)}$ and $\Bt^{(m+1)}$ is less than some specified stopping criterion. At the point of convergence of this inner process, one obtains $\widetilde{\Bt}_{\Bl}$; i.e., the maximizer of the penalized spline log-likelihood function for a fixed value of $\Bl$. Once the inner process is complete, the outer process updates the smoothing parameters and the inner process is restarted. To obtain the updated smoothing parameters, say $\Bl^*=(\lambda_0^{*},...,\lambda_J^{*})^{\intercal}$, a generalized Fellner-Schall approach \citep{wood2017biometric} is adopted and makes the update as
\begin{align*}\label{update}
\lambda_j^{*2} = \frac{\text{tr}\{\BS^{-}\BS_j\}-\text{tr}\left\{\mathcal{I}_{n,\Bl}(\widetilde{\Bt}_\Bl)^{-1} \BS_j\right\}}{{\widetilde{\Bt}_\Bl}^{\intercal} \BS_j \widetilde{\Bt}_\Bl }\lambda_j^2,\text{ for } 0\le j\le J,
\end{align*}
where $\BS^{-}$ is the generalized inverse of $\BS$. Using the updated smoothing parameters (i.e., $\Bl^*$), the inner process is then utilized to identify $\widetilde{\Bt}_{\Bl^*}$. This nested iterative process continues until the difference between $\widetilde{\Bt}_{\Bl}$ and $\widetilde{\Bt}_{\Bl^*}$ is less than some specified stopping criterion. It is worth while to point out that the update of $\Bl_j$ is guaranteed to be positive, which is required, as long as $\mathcal{I}_{n,\Bl}(\widetilde{\Bt}_\Bl)$ is positive definite \citep{wood2017biometric}, which is the case for the proposed model. At convergence of the proposed hybrid algorithm, let $\widetilde{\Bt}=(\widetilde{\Bb}^\intercal,\widetilde{\Bg}^\intercal, \widetilde{\Ba}^\intercal)^\intercal$ denote the final update of $\Bt$ with $\widetilde{\eta}(\cdot)=\eta_n(\cdot~\mid \widetilde{\gamma})$ and $\widetilde{\varphi}_j(\cdot) = \varphi_{nj}(\cdot~\mid \widetilde{\alpha}_j)$ being the proposed penalized spline estimators of $\eta(\cdot)$ and $\varphi_j(\cdot)$, respectively. 

\subsection{Variance estimation}\label{section:variance}

For the purposes of conducting large-sample inference an estimator of the variance-covariance matrix of $\widetilde{\Bb}$ was developed. This estimator is given by the upper $q\times q$ submatrix of
$$
\Sigma(\widetilde{\Bt})=(\mathcal{X}^\intercal\Omega(\widetilde{\Bt})\mathcal{X}+S)^{-1}\mathcal{X}^\intercal\Omega(\widetilde{\Bt})\mathcal{X}(\mathcal{X}^\intercal\Omega(\widetilde{\Bt})\mathcal{X}+S)^{-1}.
$$
By applying the least-square based variance estimation approach \citep{Zhang2010spline}, it can be shown that the proposed estimator consistently estimates $\mathscr{I}^{-1}(\Bb_0)$. Using this estimator one can conduct standard Wald type inference about the regression coefficients.


\section{Simulation Study}
\noindent To assess the finite sample performance of the proposed methodology the following simulation study was conducted. In this study, the true failure time $T$ was generated  according to the following model
\begin{equation}\label{model_sim}
\tg_\alpha\{F(t\mid \BX)\} =Z_1\beta_1 + Z_2\beta_2+ \eta(t) + \varphi_1(W_1)+ \varphi_2(W_2),
\end{equation}
where $W_j \sim \mathcal{U}[-1,1]$,  $Z_1 \sim \text{Bernoulli}(0.5)$ and $Z_2\sim \mathcal{N}(0,1)$. To provide for a broad range of examples, the link function $\tg_\alpha(\cdot)$ was assumed to belong to a family of links indexed by a parameter $\alpha$ as:
$$
\tg_{\alpha}(u)= \begin{cases} 
\log \Large{\frac{(1-u)^{-\alpha}-1}{\alpha}} & \text{if }\alpha > 0, \\
\log\{-\log(1-u) \}                   & \text{if }\alpha = 0.
\end{cases}
$$
Proceeding in this fashion allows us to capture several common survival models; i.e., $\alpha= 0$ leads to the PH model while $\alpha=1$ corresponds to a PO model. To examine different forms of the unknown functions and regression coefficients three scenarios were considered. In Scenario 1 (S1) the three functions were specified as $\eta(t)=\log(2t)$, $\varphi_1(w) = \exp(w+0.5)-\{\exp(1.5)-\exp(-0.5)\}/2$, and $\varphi_2(w)=2\sin(-\pi w)$ and the regression coefficients were chosen to be $\beta_1=0.5$ and $\beta_2=-0.5$; in Scenario 2 (S2) the three functions were specified as $\eta(t)=\log\{1.5t-\log(1+1.5t) \}$, $\varphi_1(w)=2\sin(-\pi w)$, and $\varphi_2(w) = 4w^2-4/3$ and the regression coefficients were set to be $\beta_1=0.5$ and $\beta_2=0.5$; in Scenario 3 (S3) the three functions were specified as $\eta(t)=\log\{\log(1+t/10)+\sqrt{t}/10\}$, $\varphi_1(w)=4w^2-4/3$, and $\varphi_2(w) = \exp(w+0.5)-\{\exp(1.5)-\exp(-0.5)\}/2$ and the regression coefficients were set to be $\beta_1=-0.5$ and $\beta_2=-0.5$. In each of S1-S3, we consider values of $\alpha\in\{0,0.5,1\}$. Note, each of the  specified functions adheres to the identifiability constraint; i.e., they integrate to zero over the support of the corresponding covariates. To create current status data, observation times ($Y$) were sampled from an exponential distribution with a mean of 2, 2, and 1 under S1, S2, and S3, respectively, and the censoring indicator was determined as $\Delta=I(T<Y)$. These specifications were made to yield a variety of right-censoring rates; for example when $\alpha=0(1)$ we have right censoring rates of  27\%(36\%), 44\%(51\%), and 77\%(81\%) under S1, S2, and S3, respectively. For each simulation configuration, this data generating process was repeated 500 times to create 500 independent data sets consisting of $n = 400$ observations, which is approximately one fourth the sample size available in our motivating data application. 

The proposed methodology was used to analyze each of the simulated data sets. For purposes of comparison, we compare the results of the proposed approach to those of the spline estimator that arises from maximizing \eqref{penalized} after dropping the penalty term. Note, this competing approach is a generalization of the technique outlined by \cite{lu2018partially} which was shown to outperform \cite{cheng2011semiparametric}. To compute both the penalized and unpenalized spline estimators, cubic $B$-spines were used to approximate all unknown functions with interior knots being placed at equally-spaced quantiles of the covariate values. For each data set, the proposed approach made use of a single knot set consisting of $\lceil n^{1/3}\rceil$ interior knots, while the unpenalized estimator was computed using multiple knot set configurations with the number of interior knots ranging from $\lceil n^{1/3} \rceil - 3$ to $\lceil n^{1/3} \rceil + 3$, with the final unpenailzed estimator being selected based on the BIC criterion; for further discussion see \cite{lu2018partially}.

Table \ref{tab:simu1} summarizes the results of estimating the regression coefficients across all considered simulation configurations for both the penalized and unpenalized spline estimator. Provided in this summary are the empirical bias, sample standard deviation, and mean squared error of the 500 point estimates, as well as the average of the 500 estimated standard errors and empirical coverage  probability associated with 95\% confidence interval. From these results it is apparent that the proposed approach works well; i.e., the penalized spline estimators of the regression coefficients exhibit little bias, the standard deviations of the point estimates are in agreement with the average estimated standard errors, and the empirical coverage probabilities are at their nominal level. The latter two findings tend to suggest that the proposed method of estimating the variance-covariance matrix works well and that large-sample inference is possible even for relatively small sample sizes. Further, the proposed penalized spline estimator outperformed in virtually every aspect its unpenalized counterpart even though this competing estimator was chosen as the best from amongst multiple candidates. Moreover, the performance of this competing technique varied dramatically across the different knot set configurations; i.e., the unpenalized spline estimator was found to be sensitive to the specification of the number and placement of the basis functions. However, penalized estimator was robust to the specification of the basis functions.    

\begin{figure}
	\includegraphics[width=1\linewidth]{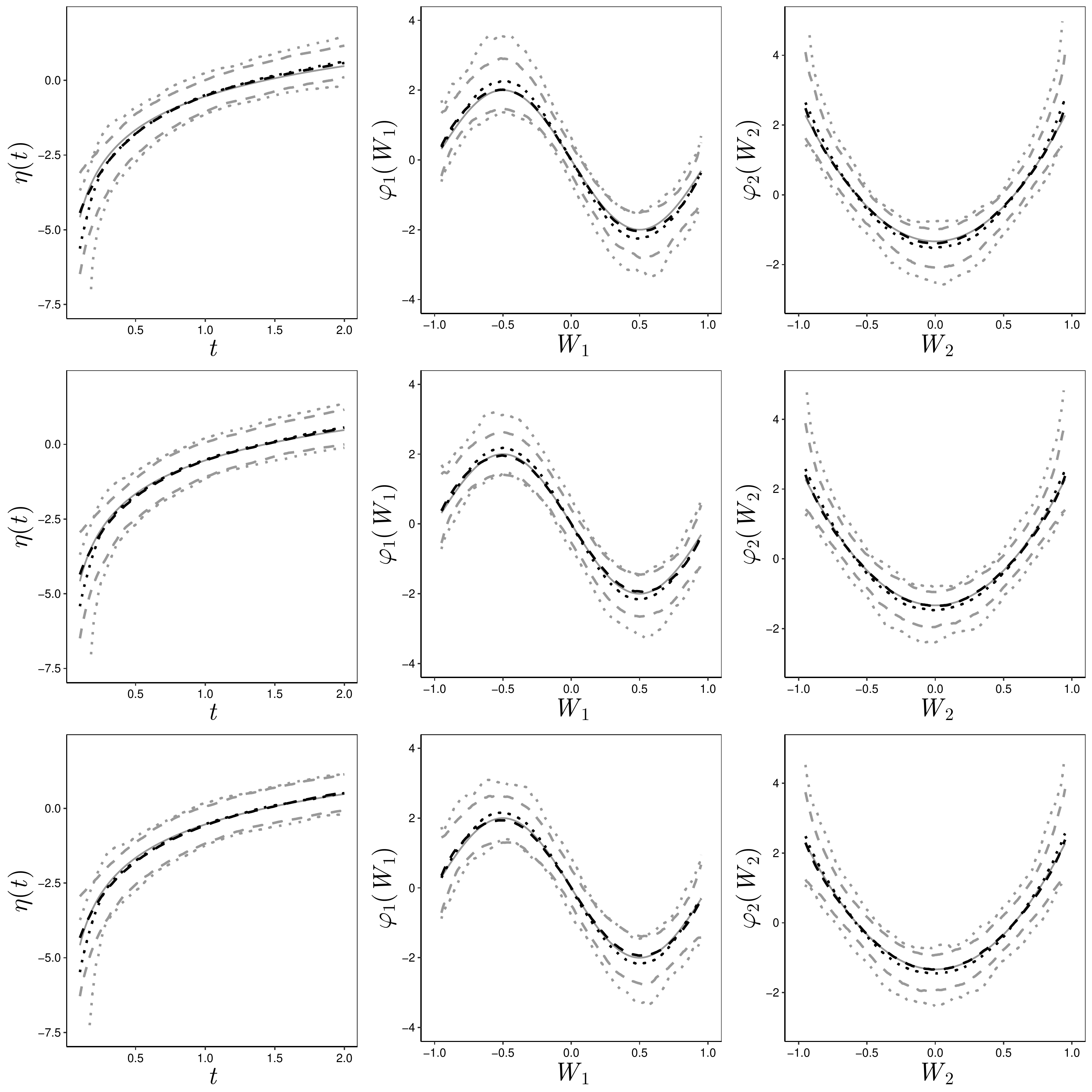}
	\caption{Estimates of the unknown functions under scenario S2 for all three models (from top to bottom: $\alpha=0, 0.5, 1.0$). The solid grey curves correspond to the true value of the functions. Dashed lines are the medians (black) and 95\% percentiles (grey) of the estimates obtained from the penalized spline approach. Dotted curves are the medians (black) and 95\% percentiles (grey) of the estimates obtained from the unpenalized spline approach.}
	\label{fig:simu1}
\end{figure}

One of the primary goals of this study was to assess the performance of the proposed method with respect to estimating the unknown functions. To this end, Figure \ref{fig:simu1} summarizes the estimates of the unknown functions for S2 for all considered values of $\alpha$ that were obtained from both approaches. This summary includes the 0.025, 0.5, and 0.975 point-wise quantiles of the estimates. The analogous figures under S1 and S3 can be found in Appendix 2 of the Supplementary Material. These results suggest that the proposed methodology can be used to accurately estimate all of the unknown functions; i.e. very little bias can be observed when one compares the the median of the functional estimates to the true underlying function. Additionally, these findings demonstrate that the penalized spline estimator results in functional estimates that have smaller variability when compared to the unpenalized spline estimator. These finding reinforce the assertion that the proposed approach performs better with respect to both estimation and inference than its unpenalized counterpart, and moreover, that it is robust to the specification of the spline functions.

\section{Application to Uterine leiomyomata data}

For further illustration, the proposed methodology is now used to analyze uterine leiomyomata (fibroid) data collected as a part of a prospective cohort study concerned with early pregnancy. Uterine fibroids are noncancerous growths that appear during childbearing years and have been found to be associated with complications in pregnancy and delivery \citep{coronado2000complications}. As a part of this study, a patient's fibroid status was determined during an ultrasound examination; i.e., during this examination fibroids were either found to be present or absent indicating that the fibroid onset time was smaller or larger than the examination time, respectively. Thus, the fibroid onset time is either left- or right-censored with the patient's age at examination being the censoring time. Of the $n = 1604$ participants considered in this analysis 216 were found to be fibroid positive at the time of their examination, leading to a left-censoring rate of approximately 13\%.

In this analysis, we focus on four risk factors purported to be related to fibroid development; i.e., race ($Z_1$ = 1 if African American and $Z_1$ = 0 Caucasian), parity status ($Z_2$ = 1 if they had given birth before and $Z_2$ = 0 otherwise), obesity status ($Z_3$ = 1 indicating obesity and $Z_3$ = 0 otherwise), and standardized age of menarche ($W$). To relate these risk factors to the fibroid onset time, the following model was specified 
\begin{equation*}
\tg\{F(t\mid \BX)\} = Z_1\beta_1 + Z_2\beta_2+ Z_3\beta_3+\eta(t) + \varphi(W),
\end{equation*}
where $\tg(u)=\log\{-\log(1-u)\}$. This link function leads to the usual PH model. Cubic $B$-splines with a knot set consisting of $\lceil n ^{1/3}\rceil=12$ interior knots were used to approximate all unknown functions. Model fitting was completed using the hybrid algorithm outlined in Section 4. The estimator of the variance-covariance matrix of $\widetilde{\Bb}$ developed in Section 4 was used to provide for standard errors and $95\%$ Wald confidence intervals for the estimated regression coefficients.  

\begin{figure}[!htb]
	\centering
	\begin{minipage}{.5\textwidth}
		\centering
		\includegraphics[width=1\linewidth]{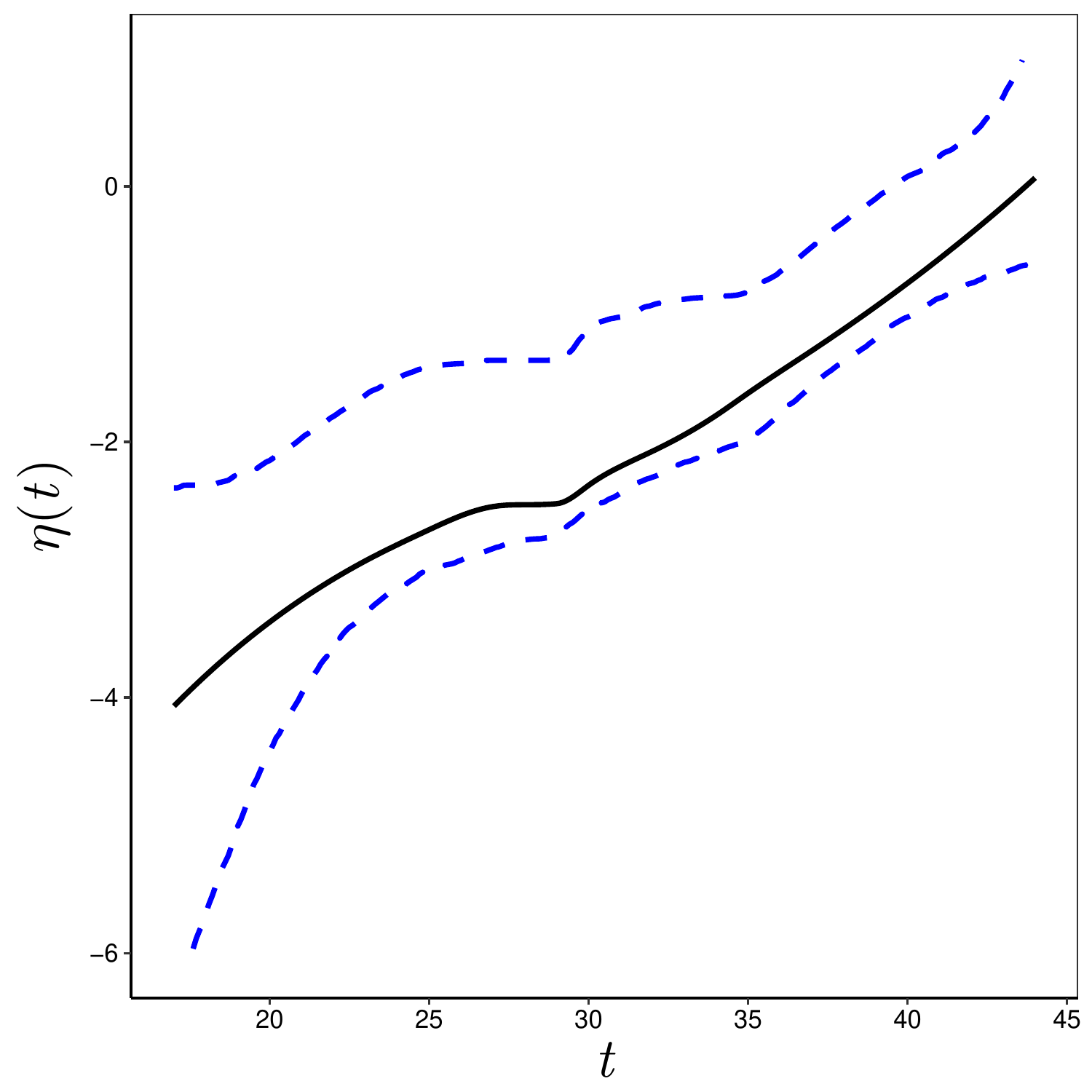}
	\end{minipage}%
	\begin{minipage}{0.51\textwidth}
		\centering
		\includegraphics[width=1\linewidth]{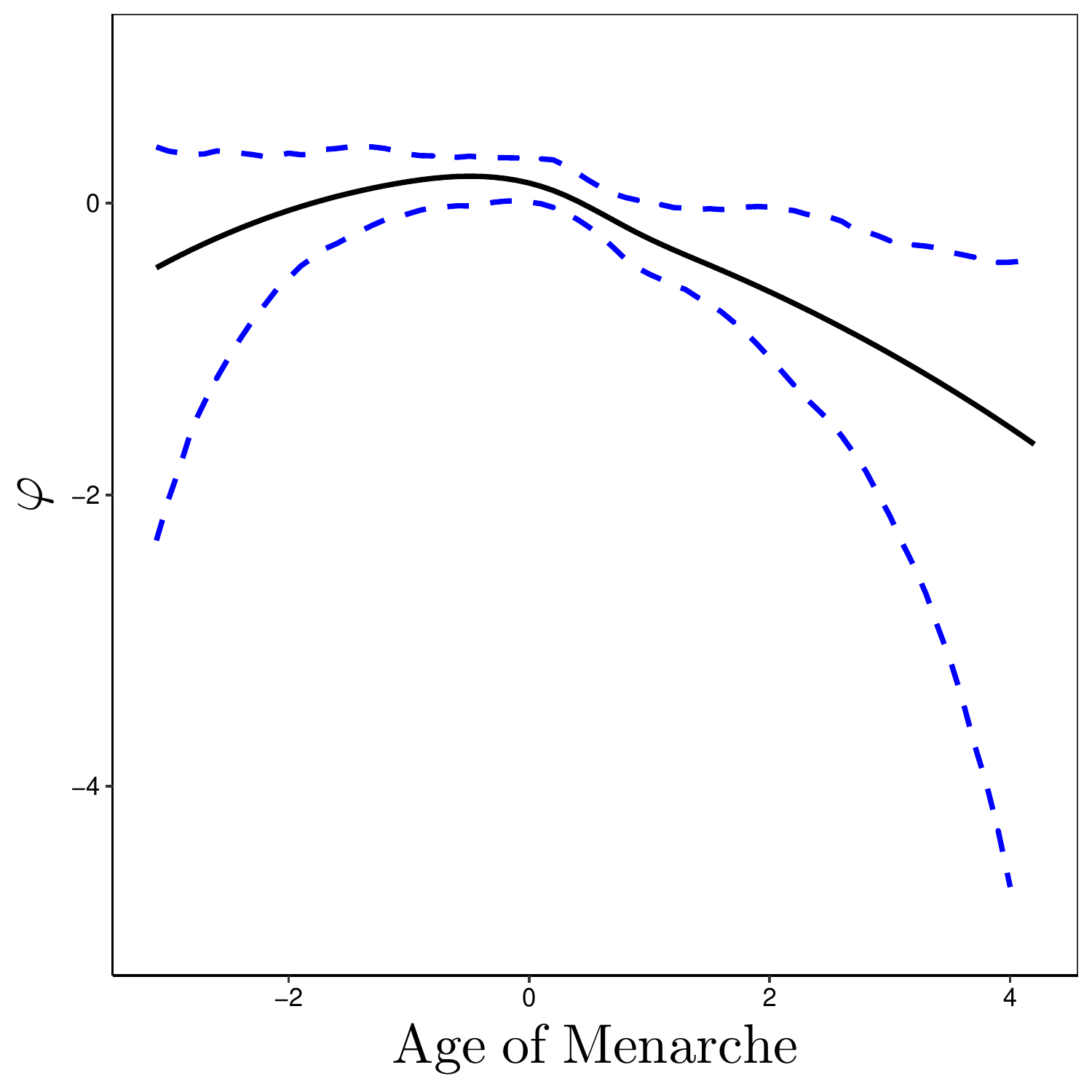}
	\end{minipage}
	\caption{Penalized functional estimate for $\eta(\cdot)$ and standardized age of Menarche under PH model. Solid black curve corresponds to estimated function based on the collected dataset, dashed curves correspond to the 2.5\% and 97.5\% percentiles of 1000 estimated functions using bootstrap samples.}\label{fig:PH_real_1}
\end{figure}

Table \ref{tab:ana_real_1} summarizes the regression parameter estimates, along with their estimated standard errors and corresponding 95\% confidence intervals. The results suggest that race and parity are significant factors, while obesity is not. In particular, these results suggest that African American women have a higher hazard of developing fibroids when compared to Caucasian women, likewise it appears that women who have not given birth before appeared to have a higher hazard of developing fibroids when compared to those that had. These results and conclusions reinforce those presented in \cite{wang2011semiparametric} and \cite{lu2018partially}. Figure \ref{fig:PH_real_1} provides the estimates of $\eta(\cdot)$ and $\varphi(\cdot)$, along with 95\% point-wise confidence intervals. The upper and lower limits of the 95\% point-wise confidence intervals were obtained as the 0.025 and 0.975 point-wise quantiles of 1000 bootstrap estimates. The estimate of $\varphi(\cdot)$ clearly exhibits a nonlinear pattern; i.e., standardized age of menarche has a nonlinear impact on the development of fibroids. To examine the effect of the assumed model form, the aforementioned analysis was conducted under the PO model by specifying $\tg(x) = \log\{x/(1-x)\}$. The estimates of the regression  coefficients under the PO model are presented in Table \ref{tab:ana_real_1}. These results provide no contradiction to the results obtained under PH model. A figure summarizing the estimates of $\eta(\cdot)$ and $\varphi(\cdot)$ under the PO model, along with 95\% point-wise confidence intervals, is provided in Appendix 2 of the Supplementary Material, and these estimated functions exhibit similar features to those obtained under PH model.

\section{Summary and discussion}
In this work we developed a computationally efficient penalized approach for fitting flexible partially linear additive transformation models to current status data. The proposed procedure demonstrates appealing properties, both numerical and theoretical. The hybrid algorithm developed for model fitting is easy to implement and robust to initialization. Through the use of penalization, the proposed approach attains superior finite-sample performance when compared to the standard spline alternative. A simple variance-covariance estimation procedure was established and was shown to provide reliable Wald-type inference. In addition to attractive finite sample performance, the estimators of the nonparametric components are shown to attain the optimal rate of convergence and we prove that estimators of regression coefficients are asymptotically normal and efficient in the semi-parametric sense.  

Given the successes of this approach, future work will be directed at generalizing this methodology to allow for the analysis of interval-censored data. A key obstacle in the development of this methodology will be the model fitting strategy, which will be far more complicated due to the complex nature of interval-censored data. Another direction for future research could involve further extending this work to allow for multiple end points, both for current status and interval censored data. Lastly, and potentially most challenging, would be the development of goodness-of-fit tests that could be used to guide the specification of the link function. This is a notoriously hard problem in the transformation model setting, and is only made harder when one considers censored data, and/or additive models.

\section*{Acknowledgement}
This research was supported by the National Institutes of Health grant AI121351 and National Science Foundation grant OIA-1826715.

\section*{Supplementary material}
\label{SM}
Technical details and proofs of Theorems 1 and 2 are provided in the Supplementary Materials, along with the additional results referenced in Sections 5 and 6.

\bibliographystyle{apalike}
\bibliography{paper-ref}


\newpage

\begin{table}
	\caption{Summary of the results that were obtained using the spline estimator and the penalized spline estimator when $\alpha=0$ (PH model), $\alpha=0.5$, and $\alpha=1$ (PO model). Presented results include the empirical bias (Bias), sample standard deviation (SD) and mean squared error (MSE) of the 500 point estimates, as well as the average of the 500 estimated standard errors (SE) and empirical coverage  probabilities associated with 95\% confidence intervals (CP)}{%
		\centering 
		\begin{tabular}{llrrrrlrrrrl} \\[-2ex]
			&  &\multicolumn{5}{c}{Spline approach} &\multicolumn{5}{c}{Penalized approach} \\  [1.0ex]
			$\alpha=0.0$  & &   Bias  & SD  &MSE&SE &CP&Bias  &SD &MSE &SE &CP   \\  
			S1
			&$\beta_1$&0.065 &0.295 &0.092 &0.252 &91.4 &-0.012 &0.234 &0.055 &0.225 &93.6 \\ 
			&$\beta_2$&-0.078 &0.174 &0.036 &0.139 &88.0 &-0.014 &0.129 &0.017 &0.122 &94.2 \\[0.5ex]
			S2
			&$\beta_1$&0.070 &0.288 &0.088 &0.252 &91.6 &0.022 &0.250 &0.063 &0.232 &94.8\\
			&$\beta_2$&0.071 &0.162 &0.031 &0.137 &90.0 &0.028 &0.140 &0.020 &0.125 &93.4 \\[0.5ex] 
			S3
			&$\beta_1$&-0.064 &0.327 &0.111 &0.284 &91.8 &-0.030 &0.292 &0.086 &0.259 &93.4\\
			&$\beta_2$&-0.071 &0.177 &0.036 &0.149 &90.6 &-0.034 &0.158 &0.026 &0.136 &92.0 \\[0.5ex] \\
			$\alpha=0.5$  & &   Bias  & SD  &MSE&SE &CP&Bias  &SD &MSE &SE &CP   \\
			S1
			&$\beta_1$&0.084 &0.303 &0.099 &0.278 &93.4 &0.023 &0.268 &0.072 &0.257 &93.8 \\ 
			&$\beta_2$&-0.060 &0.166 &0.031 &0.147 &91.8 &-0.013 &0.145 &0.021 &0.135 &93.6 \\[0.5ex]
			S2
			&$\beta_1$&0.049 &0.304 &0.095 &0.283 &92.4 &0.006 &0.275 &0.075 &0.265 &93.4\\
			&$\beta_2$&0.047 &0.171 &0.031 &0.150 &92.8 &0.010 &0.155 &0.024 &0.139 &93.4 \\[0.5ex] 
			S3
			&$\beta_1$&-0.073 &0.356 &0.132 &0.324 &93.0 &-0.041 &0.325 &0.107 &0.298 &93.2\\
			&$\beta_2$&-0.061 &0.177 &0.035 &0.169 &94.4 &-0.025 &0.159 &0.026 &0.155 &94.4  \\[0.5ex] \\
			$\alpha=1.0$  & &   Bias  & SD  &MSE&SE &CP&Bias  &SD &MSE &SE &CP   \\
			S1
			&$\beta_1$&0.036 &0.326 &0.107 &0.306 &93.4 &-0.013 &0.294 &0.087 &0.288 &94.2 \\ 
			&$\beta_2$&-0.050 &0.158 &0.027 &0.159 &95.4 &-0.015 &0.144 &0.021 &0.149 &96.6 \\[0.5ex]
			S2
			&$\beta_1$&0.073 &0.328 &0.113 &0.315 &93.6 &0.032 &0.298 &0.090 &0.297 &95.2\\
			&$\beta_2$&0.041 &0.176 &0.033 &0.164 &93.8 &0.007 &0.164 &0.027 &0.154 &93.6 \\[0.5ex] 
			S3
			&$\beta_1$&-0.035 &0.393 &0.156 &0.363 &94.2 &-0.007 &0.351 &0.123 &0.333 &93.4\\
			&$\beta_2$&-0.064 &0.194 &0.042 &0.188 &93.8 &-0.021 &0.173 &0.030 &0.172 &95.8  
	\end{tabular}}
	\label{tab:simu1}
\end{table}


\begin{table}
	\caption{Uterine fibroid data: Presented results include the parameter estimates, estimated standard errors (SE) and 95\% Wald type confidence intervals (95\% C.I.). The results were obtained using the penalized spline estimator under both the PH and PO model}{%
		\centering 
		\begin{tabular}{llrrr}  \\
			\textbf{Model}  & Parameter &   Estimate & SE  & 95\% C.I.     \\  [1.5ex]
			\textbf{PH} & Race & 1.479 & 0.165 & (1.154,  1.803)  \\
			& Parity & -0.305& 0.144 & (-0.588, -0.022)   \\
			& Obesity & 0.114 & 0.174 & (-0.227,  0.455)  \\[1.0ex]
			\textbf{PO} & Race & 1.648 & 0.200 & (1.256,  2.040)  \\
			& Parity & -0.341& 0.163 & (-0.661 -0.022)   \\
			& Obesity & 0.155 & 0.200 & (-0.237  0.547)   
	\end{tabular}}
	\label{tab:ana_real_1}
\end{table}

\end{document}